\documentclass{article}

\usepackage{PRIMEarxiv}

\usepackage[utf8]{inputenc} 
\usepackage[T1]{fontenc}    
\usepackage{hyperref}       
\usepackage{url}            
\usepackage{booktabs}       
\usepackage{amsfonts}       
\usepackage{nicefrac}       
\usepackage{microtype}      
\usepackage{lipsum}
\usepackage{fancyhdr}       
\usepackage{graphicx}       
\graphicspath{{media/}}     

\usepackage{hyperref, url}
\usepackage{listings}
\usepackage{xcolor}

\definecolor{verylightgray}{rgb}{.97,.97,.97}

\lstdefinelanguage{Solidity}{
	keywords=[1]{anonymous, assembly, assert, balance, break, call, callcode, case, catch, class, constant, continue, constructor, contract, debugger, default, delegatecall, delete, do, else, emit, event, experimental, export, external, false, finally, for, function, gas, if, implements, import, in, indexed, instanceof, interface, internal, is, length, library, log0, log1, log2, log3, log4, memory, modifier, new, payable, pragma, private, protected, public, pure, push, require, return, returns, revert, selfdestruct, send, solidity, storage, struct, suicide, super, switch, then, this, throw, transfer, true, try, typeof, using, value, view, while, with, addmod, ecrecover, keccak256, mulmod, ripemd160, sha256, sha3}, 
	keywordstyle=[1]\color{blue}\bfseries,
	keywords=[2]{address, bool, byte, bytes, bytes1, bytes2, bytes3, bytes4, bytes5, bytes6, bytes7, bytes8, bytes9, bytes10, bytes11, bytes12, bytes13, bytes14, bytes15, bytes16, bytes17, bytes18, bytes19, bytes20, bytes21, bytes22, bytes23, bytes24, bytes25, bytes26, bytes27, bytes28, bytes29, bytes30, bytes31, bytes32, enum, int, int8, int16, int24, int32, int40, int48, int56, int64, int72, int80, int88, int96, int104, int112, int120, int128, int136, int144, int152, int160, int168, int176, int184, int192, int200, int208, int216, int224, int232, int240, int248, int256, mapping, string, uint, uint8, uint16, uint24, uint32, uint40, uint48, uint56, uint64, uint72, uint80, uint88, uint96, uint104, uint112, uint120, uint128, uint136, uint144, uint152, uint160, uint168, uint176, uint184, uint192, uint200, uint208, uint216, uint224, uint232, uint240, uint248, uint256, var, void, ether, finney, szabo, wei, days, hours, minutes, seconds, weeks, years},	
	keywordstyle=[2]\color{teal}\bfseries,
	keywords=[3]{block, blockhash, coinbase, difficulty, gaslimit, number, timestamp, msg, data, gas, sender, sig, value, now, tx, gasprice, origin},	
	keywordstyle=[3]\color{violet}\bfseries,
	identifierstyle=\color{black},
	sensitive=false,
	comment=[l]{//},
	morecomment=[s]{/*}{*/},
	commentstyle=\color{gray}\ttfamily,
	stringstyle=\color{red}\ttfamily,
	morestring=[b]',
	morestring=[b]"
}

\definecolor{dkgreen}{rgb}{0,0.6,0}
\definecolor{ltblue}{rgb}{0,0.4,0.4}
\definecolor{dkviolet}{rgb}{0.3,0,0.5}
\definecolor{dkblue}{rgb}{0.0, 0.0, 0.55}
\definecolor{dkred}{rgb}{0.55, 0.0, 0.0}

\lstdefinelanguage{Coq}{ 
    mathescape=true,
    texcl=false, 
    morekeywords=[1]{Section, Module, End, Require, Import, Export,
        Variable, Variables, Parameter, Parameters, Axiom, Hypothesis,
        Hypotheses, Notation, Local, Tactic, Reserved, Scope, Open, Close,
        Bind, Delimit, Definition, Let, Ltac, Fixpoint, CoFixpoint, Add,
        Morphism, Relation, Implicit, Arguments, Unset, Contextual,
        Strict, Prenex, Implicits, Inductive, CoInductive, Record,
        Structure, Canonical, Coercion, Context, Class, Global, Instance,
        Program, Infix, Conjecture, Theorem, Lemma, Corollary, Proposition, Fact,
        Remark, Example, Proof, Goal, Save, Qed, Defined, Hint, Resolve,
        Rewrite, View, Search, Show, Print, Printing, All, Eval, Check,
        Projections, inside, outside, Def, Abort},
    morekeywords=[2]{forall, exists, exists2, fun, fix, cofix, struct,
        match, with, end, as, in, return, let, if, is, then, else, for, of,
        nosimpl, when},
    morekeywords=[3]{Type, Prop, Set, true, false, option},
    morekeywords=[4]{pose, set, move, case, elim, apply, clear, hnf,
        intro, intros, generalize, rename, pattern, after, destruct,
        induction, using, refine, inversion, injection, rewrite, congr,
        unlock, compute, ring, field, fourier, replace, fold, unfold,
        change, cutrewrite, simpl, have, suff, wlog, suffices, without,
        loss, nat_norm, assert, cut, trivial, revert, bool_congr, nat_congr,
        symmetry, transitivity, auto, split, left, right, autorewrite},
    morekeywords=[5]{by, done, exact, reflexivity, tauto, romega, omega,
        assumption, solve, contradiction, discriminate},
    morekeywords=[6]{do, last, first, try, idtac, repeat},
    morecomment=[s]{(*}{*)},
    showstringspaces=false,
    morestring=[b]",
    morestring=[d],
    tabsize=3,
    extendedchars=false,
    sensitive=true,
    breaklines=false,
    basicstyle=\small,
    captionpos=b,
    columns=[l]flexible,
    identifierstyle={\ttfamily\color{black}},
    keywordstyle=[1]{\ttfamily\color{dkviolet}},
    keywordstyle=[2]{\ttfamily\color{dkgreen}},
    keywordstyle=[3]{\ttfamily\color{ltblue}},
    keywordstyle=[4]{\ttfamily\color{dkblue}},
    keywordstyle=[5]{\ttfamily\color{dkred}},
    stringstyle=\ttfamily,
    commentstyle={\ttfamily\color{dkgreen}},
    literate=
    {\\forall}{{\color{dkgreen}{$\forall\;$}}}1
    {\\exists}{{$\exists\;$}}1
    {<-}{{$\leftarrow\;$}}1
    {=>}{{$\Rightarrow\;$}}1
    {==}{{\code{==}\;}}1
    {==>}{{\code{==>}\;}}1
    {->}{{$\rightarrow\;$}}1
    {<->}{{$\leftrightarrow\;$}}1
    {<==}{{$\leq\;$}}1
    {\#}{{$^\star$}}1 
    {\\o}{{$\circ\;$}}1 
    {\@}{{$\cdot$}}1 
    {\/\\}{{$\wedge\;$}}1
    {\\\/}{{$\vee\;$}}1
    {++}{{\code{++}}}1
    {~}{{\ }}1
    {\@\@}{{$@$}}1
    {\\mapsto}{{$\mapsto\;$}}1
    {\\hline}{{\rule{\linewidth}{0.5pt}}}1
}[keywords,comments,strings]

\title{A Blockchain-Based Approach for Collaborative Formalization of Mathematics and Programs}

\author{
  Jin Xing Lim \\
  Singapore University of Technology and Design \\
  \texttt{jinxing\_lim@mymail.sutd.edu.sg} \\
   \And
  Barnab\'e Monnot \\
  Ethereum Foundation \\
  \texttt{barnabe.monnot@ethereum.org} \\
  \AND
  Shaowei Lin \\
  \texttt{shaowei@gmail.com} \\
  \And
  Georgios Piliouras \\
  Singapore University of Technology and Design \\
  \texttt{georgios@sutd.edu.sg} \\
}

\begin{document}
\maketitle

\begin{abstract}
Formalization of mathematics is the process of digitizing mathematical knowledge, which allows for formal proof verification as well as efficient semantic searches. Given the large and ever-increasing gap between the set of formalized and unformalized mathematical knowledge, there is a clear need to encourage more computer scientists and mathematicians to solve and formalize mathematical problems together. With blockchain technology, we are able to decentralize this process, provide time-stamped verification of authorship and encourage collaboration through implementation of incentive mechanisms via smart contracts. Currently, the formalization of mathematics is done through the use of proof assistants, which can be used to verify programs and protocols as well. Furthermore, with the advancement in artificial intelligence (AI), particularly machine learning, we can apply automated AI reasoning tools in these proof assistants and (at least partially) automate the process of synthesizing proofs. In our paper, we demonstrate a blockchain-based system for collaborative formalization of mathematics and programs incorporating both human labour as well as automated AI tools. We explain how Token-Curated Registries (TCR) and smart contracts are used to ensure appropriate documents are recorded and encourage collaboration through implementation of incentive mechanisms respectively. Using an illustrative example, we show how formalized proofs of different sorting algorithms can be produced collaboratively in our proposed blockchain system.
\end{abstract}

\keywords{
blockchain \and formalized mathematics \and verifiable programs \and token-curated registry \and smart contracts \and human-AI collaboration
}


\section{Introduction} \label{sec:intro}

Formalization of mathematics is the process of digitizing mathematical knowledge into machine code so that proofs can be formally and automatically checked and recorded within a computer system. This idea was initially proposed in the \textit{QED manifesto} \cite{boyer1994qed} during the 1994 CADE conference and it is currently done through the means of proof assistants such as Isabelle \cite{wenzel2008isabelle}, Coq \cite{bertot2008short} and Lean \cite{de2015lean}.
Other than verifying mathematical proofs, proof assistants can also be used to verify programs and protocols \cite{filliatre2011deductive,sergey2017programming}. With the rise of artificial intelligence, machine learning tools can be implemented to aid us in automating, and hence, formalizing proofs partially \cite{jiang2018machine}.
However, as work in this area is still at its early stage, it requires a lot of man-hours and collaboration to formalize the ever-growing mathematical knowledge and programming world. Thus, this raises our driving challenge: \textit{to design a system architecture that decentralizes this process and allows for effective collaboration between multiple intelligent agents towards formalization of mathematics and programs.}


The benefits of decentralization, time-stamped verification and smart contracts from blockchain provide us with a platform that allows fast dissemination of problems and results, verified authorship and appropriate reward allocation of partial progresses respectively.
Thus, there are various prior works, such as Qeditas \cite{whiteqeditas}, Mathcoin \cite{su2018mathcoin} and Proofgold\footnote{\url{https://proofgold.org/}}, that discuss approaches on how computer scientists and mathematicians can share their completed work, and contribute to the world of formalized mathematics through the use of blockchain. On the other hand, the \textit{Polymath project} \cite{gowers_2009}, initiated by Timothy Gowers, is a project where mathematicians collaborate to solve difficult and important mathematical problems by sharing their partial results, presented as traditional ``handwritten" proofs.
However, to our best knowledge, there is no system currently that combines these two ideas where participants (human or automated AI tools) can collaborate to solve and formalize common problems. 

To this end, we propose a blockchain-based  architecture for formalizing mathematics and programs that allows participants to share, not only completed, but also formalized partial results. Our system architecture can be broken down into three interconnected layers: \textit{data layer} (what is recorded?), \textit{client layer} (what is meaningful?) and \textit{incentive layer} (what is rewarded?). 
We will discuss how Token-Curated Registries (TCR) \cite{tcr,asgaonkar2018token} are used to filter document submissions and how smart contracts are essential to implement incentive mechanisms to encourage participants to come together and tackle common problems.
The details and underlying mechanisms for each layer of our system are described in Section \ref{sec:architecture}. Furthermore, we provide a simple but illustrative algorithmic example in Section \ref{sec:example}, where we present how different sorting algorithms can be formalized\footnote{For concreteness, we will be using Coq as our reference proof assistant, however, it can be replaced with any proof assistant with similar functionality. The Coq codes in Section \ref{sec:example} can be found in \url{https://github.com/jinxinglim/coq-chain}.} collaboratively between human and automated AI tools using our proposed blockchain-based system. 



\section{Formalized Proofs and Its Current State} \label{sec:formalproofs}

Thanks to intuitionistic type theory \cite{martin1984intuitionistic} and the Curry-Howard correspondence \cite{howard1980formulae}, mathematical knowledge can be written as code through the means of proof assistants, such as Isabelle \cite{wenzel2008isabelle}, Coq \cite{bertot2008short} and Lean \cite{de2015lean}. Such proofs are said to be \textit{formalized}. 
The main driving mechanism in proof assistants is the use of \textit{tactics} for users to guide the proof engine to derive the proofs interactively. The goal of a tactic is to break down complicated theorem statements into multiple simpler statements that are hopefully easier to complete. 

Other than having proofs formally and automatically verified by the proof assistants' proof systems, there are a lot of efforts to formalize both existing and new mathematical knowledge due to several reasons, including:


\paragraph{Verified programs and protocols} We can use proof assistants to formally verify programs, e.g., CompCert \cite{leroy2016compcert}, and protocols, e.g., CBC Casper protocol \cite{li2020formalizing}, as well. Moreover, with the extraction features, such as the one implemented in Coq \cite{filliatre2004functors}, one can even state a program specification as a theorem statement and extract its proof as a working program. 

\paragraph{Proofs with gaps} In proof assistants, we can write \textit{partial proofs} - proofs with gaps, where parts of the proofs can be left empty for future completion by either the owner or someone else. We shall use the proof of the proposition, $sum\_first\_n$, $\forall n \in \mathbb{N}, \sum_{i=1}^n k = \frac{n(n+1)}{2}$, as an illustration. The owner of Coq file shown in Listing \ref{lst:sum_first_n} created a partial proof through the use of the \lstinline[language=Coq]|induction| tactic, and left the proofs of both base and inductive cases empty as conjectures for future completion.

\begin{lstlisting}[language=Coq,numbers=none,caption=Partial proof of sum\_first\_n viewed as a Coq file, label=lst:sum_first_n,frame=single]
Conjecture base_case : sum_to 0 = 0*(0+1)/2.

Conjecture ind_case: forall (n:nat), sum_to n = n*(n+1)/2 -> sum_to (S n) = S n*(S n+1)/2.

Theorem sum_first_n : forall (n:nat), sum_to n = n*(n+1)/2.
Proof. induction n. apply base_case. apply ind_case. trivial. Qed.
\end{lstlisting}

\paragraph{Automated AI tools} There has been an increasing number of automated tools implemented for proof assistants, such as Sledgehammer \cite{meng2006automation} (for Isabelle), TacticToe \cite{gauthier2018learning} (for HOL4 \cite{slind2008brief}), CoqHammer \cite{czajka2018hammer} and Tactician \cite{blaauwbroek2020tactician} (both for Coq). These tools make use of machine learning methods such as $k$-NNs, RNNs and neural networks to help with the proof searches. The goal of most of these tools is to have an automated proof search function that generates a proof automatically given some theorem statement, and hence, ease the mathematicians' load in formalizing proofs.




\section{Related Works and Gaps} \label{sec:relatedworks}

Through the use of blockchain, decentralization enables fast dissemination of information, allowing users to be notified with first-hand problems and state-of-the-art results. If users were to collaborate to formalize and solve a common theorem statement, they do not need to worry about the authorship of their partial progresses due to the time-stamped verification provided by blockchain. Furthermore, with the use of smart contracts, one can trigger incentive mechanisms to allocate rewards appropriately to encourage users to share their partial results and collaborate to solve a common problem.

There are several prior projects such as Qeditas \cite{whiteqeditas}, Mathcoin \cite{su2018mathcoin} and Proofgold tackling the use of blockchain technology to tie sources of formalized mathematics together in a decentralized way. Nevertheless, there are several gaps that these projects do not address adequately and which we aim to improve upon:

\paragraph{Filtering submissions} Documentation of formalized mathematical objects, i.e., definitions, theorem statements and proofs, are organized in terms of libraries, such as the Mathematical Components library of Coq ssreflect \cite{gonthier2010introduction} and the Lean mathematical library, mathlib \cite{mathlib2020}. To be documented in any of these libraries, experts from the respective community will determine whether each submitted document fulfills certain criteria. Yet, in a decentralized system, there will be no ``expert'' to filter out submissions. Qeditas suggested strategies such as inclusion of ``salted'' documents and having fees proportionate to the size of the submitted documents to deter duplicate submissions. Having said that, to filter out novel, non-duplicate and relevant work from the rest, we believe a much stronger mechanism needs to be in place so that the decentralized library will be streamlined. Our approach featuring a TCR allowing users to accept or reject a given submission would be particularly useful, but has not been discussed in any of the prior works. 

\paragraph{Collaborative work} Although there are several useful ideas suggested by these prior projects to incentivize users to formalize proofs, such as rewarding ownership as intellectual property, bounties and token betting, there has been little to no mention of the use of smart contracts to implement complex incentive mechanisms to split rewards appropriately to collaborators. With the right incentive mechanism in place and the ability to leave proofs with gaps as mentioned in Section \ref{sec:formalproofs}, we aim to reward contributors for completed as well as partial results. Contributors need not worry about the authorship of their partial work due to the time-stamped verification provided by blockchain, serving as a motivation for them to submit partial, incomplete results and formalize proofs collaboratively.

\paragraph{Automated AI tools' participation} None of prior projects mention about the involvement of automated AI tools, which are discussed in Section \ref{sec:formalproofs}. 
Some of these automated tools require high computing power to produce better automated proof searches. Developers could deploy these tools as nodes in the blockchain environment to look for conjectures to solve instead of requiring users to have the necessary computing capabilities in order to use them.
Automated tools, such as TacticToe \cite{gauthier2018learning} and Tactician \cite{blaauwbroek2020tactician}, show that there is a drift towards implementing incremental proof construction \cite{paulson2020machine}, i.e., suggesting the next best proof step instead of automating the whole proof of a given theorem statement. These tools could help human provers by suggesting the ``next best move" in their proofs, facilitating collaboration between human and machines. In the future, once these automated tools reach a certain efficiency level, different tools could even come together through this blockchain environment to generate the proofs, with little to no human intervention, as an unified proof synthesizer. 




\section{System Architecture} \label{sec:architecture}

\begin{figure*}[ht]
\centering
\includegraphics[width=\linewidth]{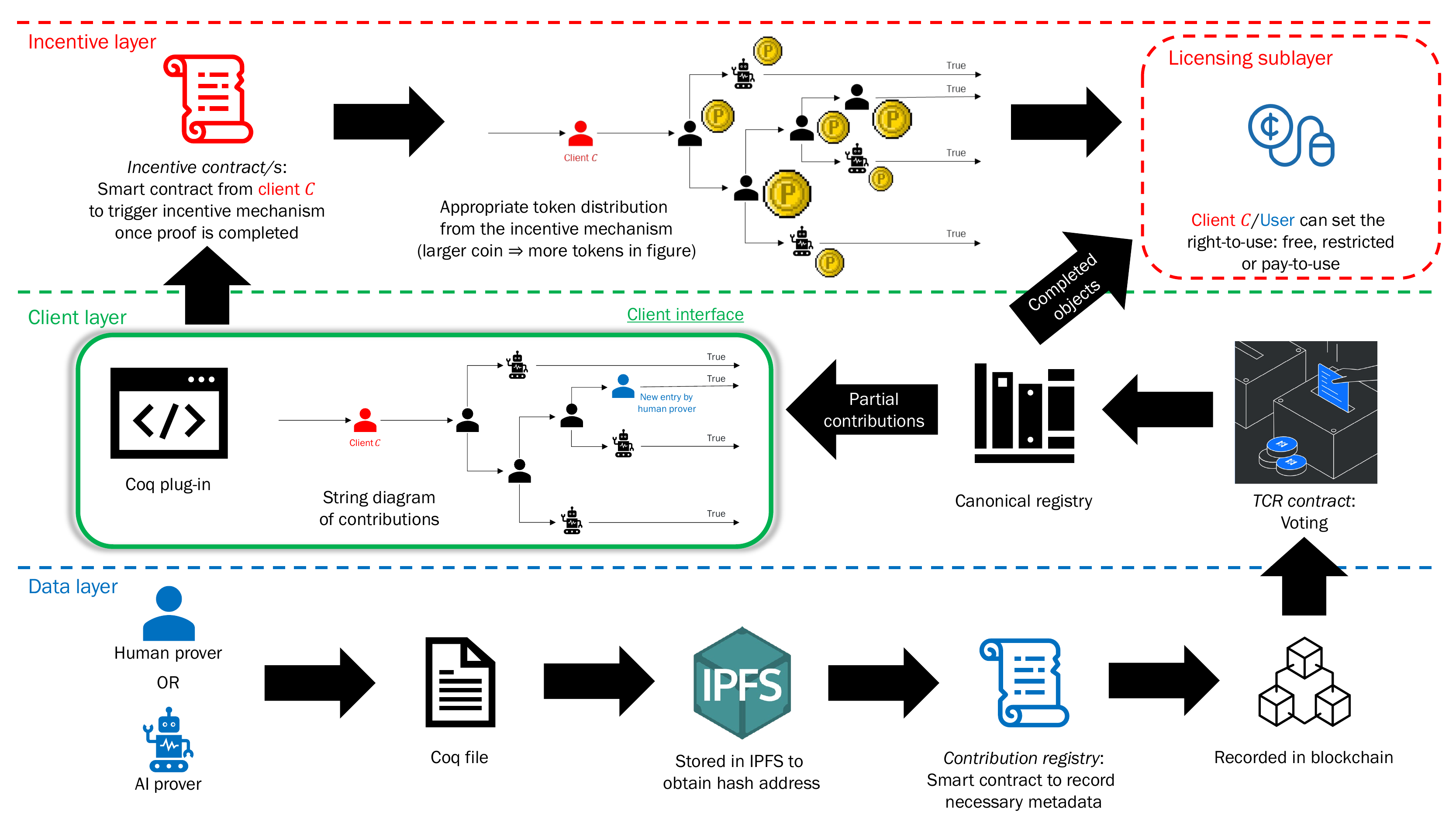}
\caption{Flowchart of how a prover's contribution gets recorded and eventually rewarded in our blockchain system across the three layers of our system architecture. Images' sources: IPFS - \url{https://en.wikipedia.org/wiki/InterPlanetary_File_System}, TCR - \url{https://0xcert.org/news/token-curated-registries-william-entriken/}, right-to-use - \url{https://reasonstreet.co/business-model-pay-per-use/}}
\label{fig:architecture_flowchart}
\end{figure*}

We assume the availability of a public, programmable blockchain, where smart contracts are deployed. We call \textit{prover} an account (or address) on the blockchain participating in the protocol. A human or AI participant (prover using automated AI tool) may operate several prover accounts. For simplicity, we consider the case where one entity corresponds to one prover only. 
Our system architecture can be broken down into three interconnected layers: 
A) the data layer (what is recorded?), B) the client layer (what is meaningful?) and C) the incentive layer (what is rewarded?).
Fig. \ref{fig:architecture_flowchart} shows a flowchart of how a prover's contribution gets recorded and eventually rewarded in our blockchain system across these three layers. The types of smart contracts and the need of bootstrapping our system with some initial library are also discussed at the end of this section. 

\subsection{Data layer}

The data layer is composed of two elements: a deployed smart contract on the blockchain, \textit{registry}, and a decentralized file \textit{storage}.

The registry accepts \textit{records} written by provers. A record contains minimal information to identify a unique \textit{contribution}. A contribution consists of a Coq file kept in the storage, while the record contains as an attribute the address of the contribution.

We suggest the use of decentralized file storage such as IPFS \cite{benet2014ipfs}. Clients running the IPFS protocol publish content (such as the Coq \texttt{.v} files) over the IPFS network. The content is addressed by a unique hash, a random string of characters, obtained by hashing the file content. Other IPFS clients interested in the content will request it by its hash to other peers on the network, until they obtain the file from a peer who is hosting it. The design pattern of hosting longer-form content, such as text, code or media over IPFS while referencing the hash in a blockchain transaction has become commonplace. While not guaranteeing availability (a file could either be missing or held offline), it is sufficient for just one party holding the file to be online for it to be available.

\subsection{Client layer} \label{subsec:client}

The data layer accepts any record and contribution. However, contributions that contain syntactically incorrect code must be disregarded by users of the protocol. The data layer cannot run the checks required to decide whether a contribution is \textit{valid} or not, as it would have to run the Coq interpreter to find out the validity of the code. While this is theoretically possible in a Turing-complete environment, the cost of such operation would be prohibitively high.

The protocol also needs to deal with the issue of spam and duplicates. Contributions of dubious quality may well be registered, as well as identical copies of previous statements, copies aggregating multiple previous contributions or copies with refactored variable names. Heuristics may be developed to ``flag'' copies by comparing their Coq types against previously published contributions, a first step towards creating a \textit{canonical registry}.

To increase the importance of \textit{novel} and \textit{useful} contributions over duplicates and tautological statements, we add an incentive for provers in the protocol to \textit{vote} on the inclusion of newly registered statements into the canonical registry. This incentive is deployed as a Token-Curated Registry (TCR) \cite{tcr,asgaonkar2018token}, a smart contract where provers stake some amount of tokens (becoming \textit{bonded provers}) to earn the right of voting in or voting out contributions from the canonical registry.

Provers may adopt the role of contributors or curators, the former when they add a record to the registry, the latter when they vote on the inclusion of a record to the canonical registry. As contributors, provers attach an inclusion stake along with their record. Any prover may challenge the inclusion of the record to the canonical registry by posting a dispute stake. As curators, provers are rewarded for voting on the correct outcome, determined by a majority rule, in the event of a challenge. If the record is voted out, the challenger receives part of the inclusion stake, while the remainder is divided among voters who rejected the record. If the record is voted in, the dispute stake of the challenger is lost to the contributor and curators who voted to include the record. Bonded provers may choose to participate in a vote or not. After some delay period, either the record enters the canonical registry unchallenged, or following a successful vote in favor of inclusion.

Although ``canonical'' appears to refer to a binary property, in effect a quantitative measure of weight distinguishes one contribution from another, as a varying number of curators engage in the voting process for each incoming record. Uncontroversial inclusion ought to bring more positive votes from curators, who need not expend a lot of time to identify novelty or usefulness, as would uncontroversial rejection, e.g., in the case of an invalid Coq code, which is also trivial to detect assuming provers run a Coq client themselves.

The client layer is responsible for presenting the current state of the canonical registry to the prover, as an interface to the data layer. Essentially, the client layer could function as a plug-in to popular Coq editors/interpreters, allowing provers to maintain and track the contributions they are working with, associated with their weights to provide contextual information on their validity, novelty and usefulness. 
The client layer additionally runs the Coq engine to decide on the validity of a contribution, as well as heuristics on the Coq types of the statement to determine their potential novelty. These checks allow the bonded validators to input their votes to the TCR via the client layer.

The client remains loosely coupled with the data layer. Any interface capable of reading data from the registry, verifying validity of the contributions and publishing progress to the registry functions as a client, allowing for diversity in implementations and features. For instance, a protocol for provers to include metadata in their contribution, e.g., as a header in the Coq file. Metadata include compatibility with Coq versions, contribution type (theorem, proof, ...), as well as imports. For imports of previous contributions, the client holds a 
string diagram (more details and examples in Section \ref{sec:example})
, linking a contribution to its ancestors. This 
string diagram
could provide the structure necessary for the definition of incentives explored in the next section. In a future iteration, it may be possible to compose the TCR instantiation described above with the contextual information received from the string diagram of proofs, akin to the citation graph of Ito and Tanaka \cite{ito2019token}. Contributions building on previously registered items would then add weight to their ancestors, indirectly recognising their usefulness and/or novelty.

\subsection{Incentive layer} \label{subsec:incentive}

The data layer and the client layer ensure consensus over the canonical registry. Consistent with our modular architecture, the incentive layer is once again loosely coupled with the previous client layer.

The incentive layer is a set of \textit{incentive mechanisms}, embodied by smart contracts. Any entity may deploy an incentive mechanism, either of their own conception or from existing templates.

We start with the simplest possible template, a fixed prize payment for the completion of a single proof. In this scenario, an entity deploys a smart contract holding the prize, with a single signer (``owner'') deciding to offer the prize to a contribution that formally proves the proposed statement. The mechanism records the identifier of the winning contribution and the author of the contribution receives the prize money.

The previous template requires trust assumptions on the signer deciding on the winner. It is easily extended to a multisignature implementation, where some set of signers must approve the winner before the prize money is unlocked. Further, it may be extended to the whole set of bonded provers.

More complex incentive systems may be added later on in a compatible manner, as long as they make reference to the original contribution 
string diagram.
For instance, staking tokens on some branch of partial progress from a statement could entitle stakers to a share of future rewards obtained should that branch terminate with a proof of the statement, with the final prover earning a larger share in proportion to the amount staked. Via this partial ownership, stakers are rewarded for detecting promising branches, while provers are induced to progress on these promising branches.



\subsubsection*{Licensing sublayer}

Up to this point, we have addressed how we can implement incentive mechanisms via smart contracts to encourage provers to collaborate and solve common open conjectures. This leads to the next question: how can we incentivize users to post new formalized definitions, theorems with completed proofs and tactics? To answer this, we will have a \textit{licensing sublayer} that allows the author to set the \textit{right-to-use} for each of his or her formalized object. This idea was introduced in one of the prior works, Qeditas \cite{whiteqeditas}, where the right-of-use can be categorized into three types of uses - \textit{free-to-use} (free to use without payment), \textit{restricted-to-use} (unable to use at all) and \textit{pay-to-use} (pay some $x$ tokens in order to use). The rights are set to free-to-use by default, and any change to this setting will have to be done by the author. We include this sublayer within the incentive layer as rewards in terms of payments can be involved to the authors of completed objects.

\subsection*{Smart Contracts}

Several types of smart contracts are available for the users in our blockchain system:

\paragraph{Contribution registry} A contract, \lstinline[language=Solidity]|registry|, stores records of contributions, including their necessary metadata such as IPFS hash address, Coq versions, contribution type (conjecture, theorem, proof, ...), imports and right-to-use. Examples of how they may look like are shown as \lstinline[language=Solidity]|ct00Cont|, \lstinline[language=Solidity]|ct01Cont| and \lstinline[language=Solidity]|ct02Cont| in Section \ref{sec:example}.

\paragraph{Incentive contracts} These contracts implement incentive mechanisms to distribute tokens to a prover or a group of provers based on their contributions and their payout rules. An example is provided as \lstinline[language=Solidity]|iCont00| in Section \ref{sec:example}.

\paragraph{TCR contract} A contract deploys TCR voting mechanism to determine the weight supporting inclusion in the canonical registry of some contribution, detailed in Section \ref{subsec:client}.

\subsection*{Bootstrapping}

The registry is easily bootstrapped with the current Coq standard library, providing the foundation for new proofs and tactics to be added to the registry. In addition, to decentralize further inputs to the TCR voting mechanism, initial allocation of voting tokens may be done with respect to the amount of work provided on the Coq library itself as well as contemporary work on proof and tactics.





\section{An Illustrative Example} \label{sec:example}

We will assume that the documents in the Coq standard library are already stored in IPFS. For simplicity, we will use \lstinline[language=Solidity]|0x0file| to represent the hash address of the Coq file, \lstinline[language=Coq]|file.v|, in IPFS.

Suppose a prover, client $C$, starts a new project to find proof/s of the specification of a sorting program as a conjecture through the Coq script, \lstinline[language=Coq]|ct00.v|, as shown in Listing \ref{lst:ct00}.

\begin{lstlisting}[language=Coq,numbers=none,caption=Coq script of \lstinline|ct00.v|, label=lst:ct00,frame=single]
Require Export Arith Sorted Permutation List.
Export List.ListNotations.
Open Scope list_scope.

Definition sorted := Sorted le.
Definition permutation := @Permutation nat.

Conjecture sort_prog : 
  forall (l:list nat), {l':list nat | sorted l' /\ permutation l' l}.
\end{lstlisting}

$C$ will upload the file in IPFS and get its hash address, \lstinline[language=Solidity]|0x0ct00|, and create a contract, \lstinline[language=Solidity]|ct00Cont|, that records the conjecture, its necessary metadata and incentive contract, \lstinline[language=Solidity]|iCont00|, as shown in Listing \ref{lst:ct00_cont}. If the proof of \lstinline[language=Coq]|sort_prog| is completed by a set of \lstinline[language=Solidity]|Provers|, \lstinline[language=Solidity]|iCont00| will run the pre-defined \lstinline[language=Solidity]|allocation| function to measure the weights of the contributions and \lstinline[language=Solidity]|allocate| the \lstinline[language=Solidity]|TokenReward| appropriately. Otherwise, all the \lstinline[language=Solidity]|TokenReward| will be transferred to the unique prover.

\begin{lstlisting}[language=Solidity,numbers=none,caption=Pseudo-code example of \lstinline|ct00Cont| and \lstinline|iCont00|, label=lst:ct00_cont,frame=shadowbox]
contract iCont00 {
    ...
    // reward to set of provers, Provers, with reward, TokenReward
    reward(list Provers) public {
        require(verify(sort_prog) == true);
        if (length(Provers) == 1) {
            transfer(Provers, TokenReward);
        }
        else {
            // weightage of rewards will based on allocation fn
            weights = allocation(sort_prog);
            allocate(Provers, weights, TokenReward);
        }
    }
    ...
}

contract ct00Cont is registry {
    ...
    file="0x0ct00";
    CoqVer="8.12";
    filetype="Conjecture";
    imports=["0x0Arith","0x0Sorted","0x0Permutation","0x0List"];
    // Activate Icont00 to set of provers = Provers
    iCont00(Provers, TokenReward);
    ...
}
\end{lstlisting}

\begin{figure}[ht]
\centering
\includegraphics[width=\linewidth]{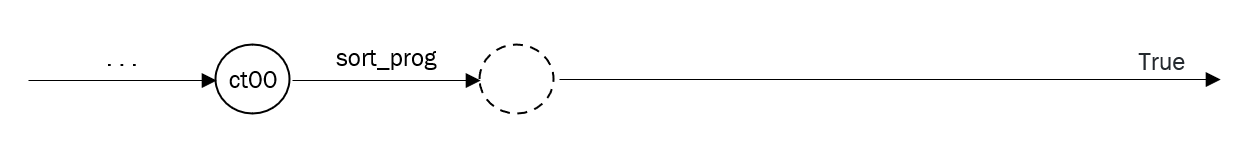}
\caption{String diagram representing partial proof of \lstinline[language=Coq]|sort_prog|. Strings (arrows) represent the theorems, solid circles represent completed subproofs and dotted circles represents incomplete subproofs (proof gaps). The goal is to complete all paths to the trivial theorem \lstinline[language=Coq]|True| from the desired theorem \lstinline[language=Coq]|sort_prog|.}
\label{fig:stringd_ct00}
\end{figure}

At the client interface, $C$ can do any validity check that cannot be done in the data layer through the Coq plug-in editor and a string diagram can be presented for $C$ to track any progression of the conjecture, as shown in Fig. \ref{fig:stringd_ct00}.

After \lstinline[language=Solidity]|ct00Cont| is posted in the blockchain system, human prover $P$ is the first one to provide a partial proof of \lstinline[language=Coq]|sort_prog| via the Coq script, \lstinline[language=Coq]|ct01.v|, as shown in Listing \ref{lst:ct01}. He initiates the proof by using induction and leaves both the base case and inductive case empty. He uploads this file in IPFS and record a \lstinline[language=Solidity]|ct01Cont| entry as shown in Listing \ref{lst:ct01_cont}.

\begin{lstlisting}[language=Coq,numbers=none,caption=Coq script of \lstinline|ct01.v|, label=lst:ct01,frame=single]
Require Export ct00.

Conjecture sort_base : {l' : list nat | sorted l' /\ permutation l' []}.

Conjecture sort_prog_IH : forall (a : nat) (l x : list nat), sorted x -> permutation x l 
    -> {l' : list nat | sorted l' /\ permutation l' (a :: l)}.

Lemma sort_prog : 
  forall (l : list nat), {l' : list nat | sorted l' /\ permutation l' l}.
Proof. induction l. apply sort_base. destruct IHl; destruct a0; eapply sort_prog_IH; eassumption. Qed.
\end{lstlisting}

\begin{lstlisting}[language=Solidity,numbers=none,caption=Pseudo-code example of \lstinline|ct01Cont|, label=lst:ct01_cont,frame=shadowbox]
contract ct01Cont is registry {
    ...
    file="0x0ct01";
    CoqVer="8.12";
    filetype="PartialProof";
    imports=["0x0ct00"];
    ...
}
\end{lstlisting}

\begin{figure}[ht]
\centering
\includegraphics[width=\linewidth]{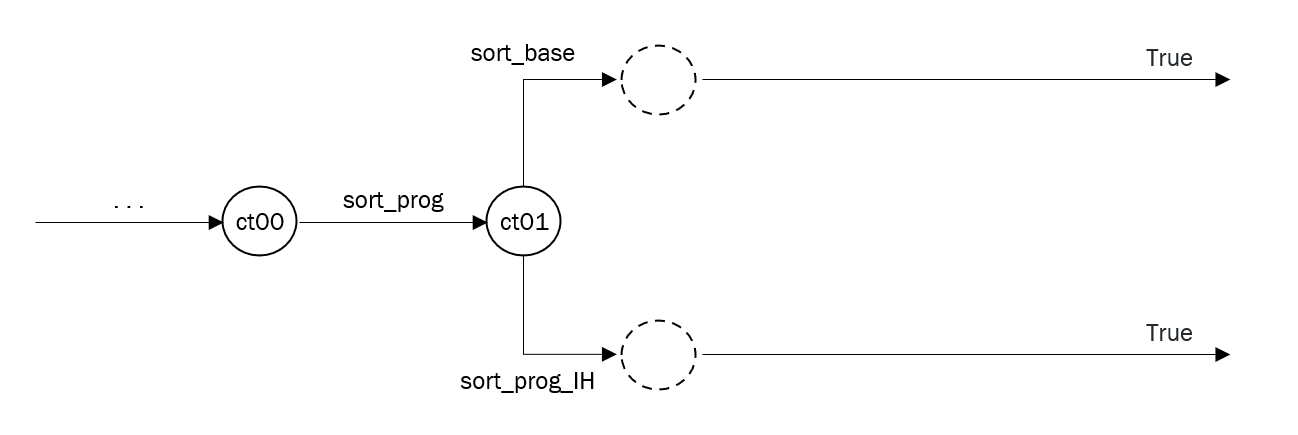}
\caption{Update in the string diagram shown in Fig. \ref{fig:stringd_ct00} after $P$ uploads \lstinline[language=Coq]|ct00.v| entry.}
\label{fig:stringd_ct01}
\end{figure}

After which, $C$ and other users will get a notification through the client interface, which shows an update in the string diagram of Fig. \ref{fig:stringd_ct00}, as shown in Fig. \ref{fig:stringd_ct01}.

\begin{figure}[ht]
\centering
\includegraphics[width=\linewidth]{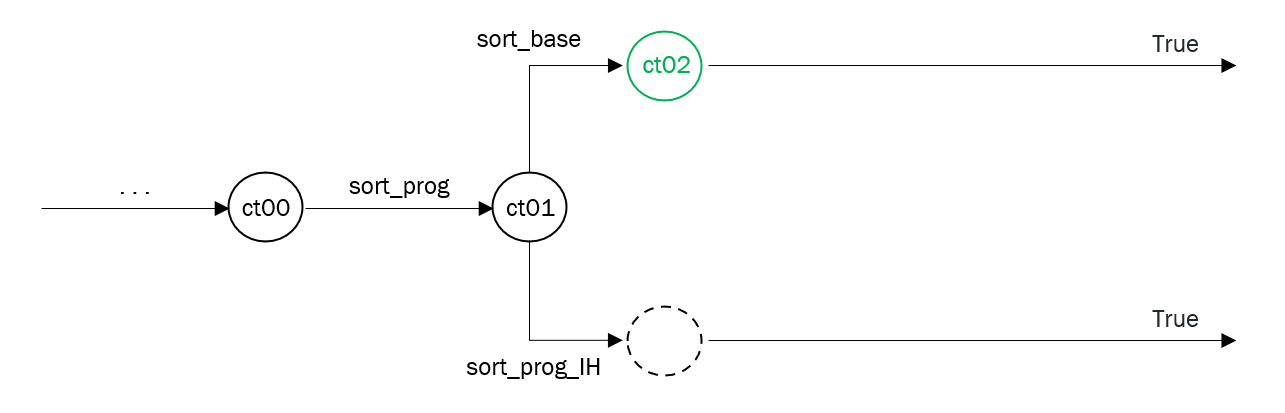}
\caption{Update in the string diagram shown in Fig. \ref{fig:stringd_ct01} after $A$ uploads \lstinline[language=Coq]|ct02.v| entry. We use green circles to represent entry from automated AI tools.}
\label{fig:stringd_ct02}
\end{figure}

At this point, the unsolved conjectures are \lstinline[language=Coq]|sort_prog|, \lstinline[language=Coq]|sort_base| and \lstinline[language=Coq]|sort_prog_IH|. An AI prover, $A$, using automated AI tool, say Coqhammer \cite{czajka2018hammer}, will try to automate the proofs for these three conjectures. Although it only manages to automate the proof of \lstinline[language=Coq]|sort_base|, it records its attempts via the Coq script, \lstinline[language=Coq]|ct02.v| (Listing \ref{lst:ct02}) and a \lstinline[language=Solidity]|ct02Cont| entry (Listing \ref{lst:ct02_cont}) in the system. The client interface will then update the progress in the string diagram shown in Fig. \ref{fig:stringd_ct01}, as shown in Fig. \ref{fig:stringd_ct02}.

\begin{lstlisting}[language=Coq,numbers=none,caption=Coq script of \lstinline|ct02.v|, label=lst:ct02,frame=single]
Require Import ct01.
From Hammer Require Import Hammer.

Lemma sort_prog : forall (l : list nat), {l' : list nat | sorted l' /\ permutation l' l}.
Proof. Fail hammer. Abort.

Lemma sort_base : {l' : list nat | sorted l' /\ permutation l' []}.
Proof. hammer. Defined.

Lemma sort_prog_IH : forall (a : nat) (l x : list nat), sorted x -> permutation x l 
    -> {l' : list nat | sorted l' /\ permutation l' (a :: l)}.
Proof. Fail hammer. Abort.
\end{lstlisting}

\begin{lstlisting}[language=Solidity,numbers=none,caption=Pseudo-code example of \lstinline|ct02Cont|, label=lst:ct02_cont,frame=shadowbox]
contract ct02Cont is registry {
    ...
    file="0x0ct02";
    CoqVer="8.12";
    filetype="CompletedProof";
    imports=["0x0ct01"];
    ...
}
\end{lstlisting}

\begin{figure}[ht]
\centering
\includegraphics[width=\linewidth]{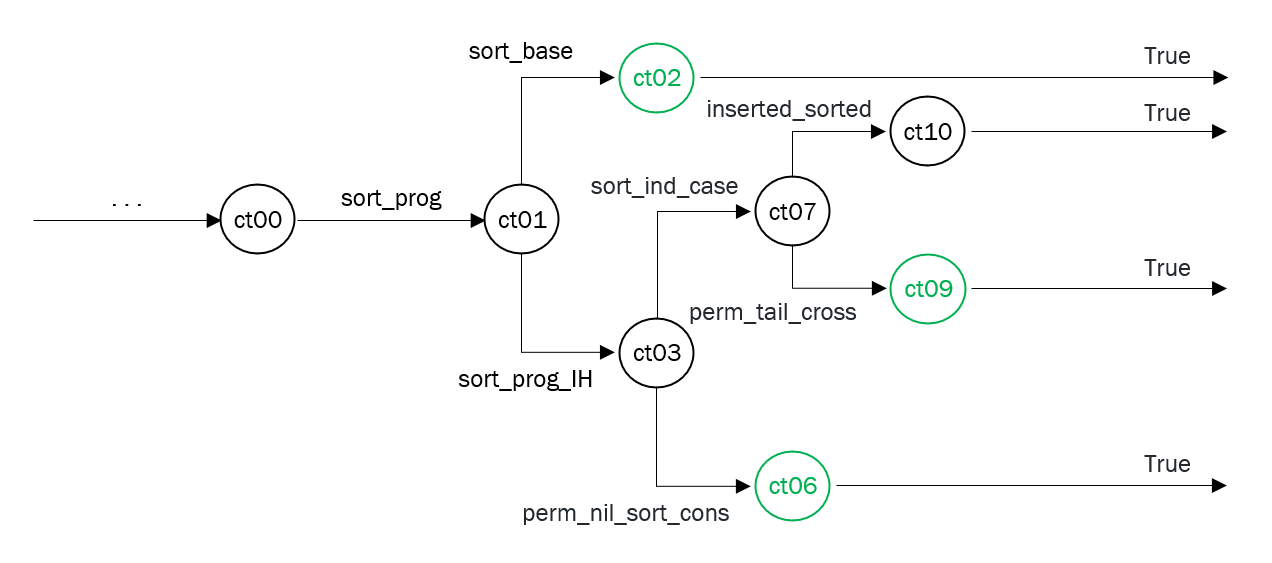}
\caption{Collaboration of contributions from human provers and automated AI tool/s that yields a proof of insertion sort algorithm.}
\label{fig:stringd_isort}
\end{figure}

Eventually all subproofs will be completed, i.e., no more dotted circles between \lstinline[language=Coq]|sort_prog| and \lstinline[language=Coq]|True|, through the collaboration of contributions from both human provers and automated AI tool/s as shown in Fig. \ref{fig:stringd_isort}. The completed proof from this network of subproofs is essentially a verification proof of an insertion sort algorithm.

Once this is achieved, the \lstinline[language=Solidity]|reward| function in \lstinline[language=Solidity]|iCont00| will be triggered to \lstinline[language=Solidity]|allocate| each contribution, seen in Fig. \ref{fig:stringd_isort}, with the appropriate rewards based on the \lstinline[language=Solidity]|allocation| function. It is important to have the right allocation rule such that it would encourage each prover to share his or her partial result immediately instead of withhold any partial progress and release only when he or she completes the whole proof. For the existence of such allocation rule, one can implement, for example, the mechanism discussed by Banerjee et al \cite{banerjee2014re}. Measuring of the weights in any allocation rule may involve the use of TCR as mentioned in Section \ref{sec:architecture}.

Besides insertion sort, there are many different sorting algorithms, including merge sort, quick sort, etc. Unlike in mathematics, where it is sufficient to have one proof for each theorem, different proofs for a specification of a program may yield algorithms with different efficiencies. Thus, it is important to implement incentive mechanism/s via smart contract/s to encourage users to produce different proofs collaboratively as well. For example, $C$ could implement an incentive mechanism similar to what was studied by Babaioff et al. \cite{babaioff2012bitcoin} via a smart contract. That is, $C$ will reward the first completed proof with \lstinline[language=Solidity]|TokenReward|, followed by the second completed proof with \lstinline[language=Solidity]|TokenReward/2|, then the third completed proof with \lstinline[language=Solidity]|TokenReward/4| and so on. This mechanism of halving the \lstinline[language=Solidity]|TokenReward| for each completed proof will eventually cost $C$ a maximum of $2 \times$\lstinline[language=Solidity]|TokenReward|. 

As a result, another prover $T$ is motivated to create an alternative proof to \lstinline[language=Coq]|sort_prog|, which will eventually lead to a proof of merge sort algorithm. This can be done by first, creating a new tactic, \lstinline[language=Coq]|div_conq_split|, which is a formalization of the divide-and-conquer algorithmic paradigm, in the Coq file \lstinline[language=Coq]|ct04.v| (Listing \ref{lst:ct04}). The file is uploaded in the IPFS network and records an \lstinline[language=Solidity]|ct04Cont| entry (Listing \ref{lst:ct04_cont}). This entry will state that anyone who wishes to use the \lstinline[language=Coq]|div_conq_split| tactic needs to pay a fee of some \lstinline[language=Solidity]|x| token/s. This is done as the tactic can be used to prove any property of lists other than the \lstinline[language=Coq]|sort_prog|.

\begin{lstlisting}[language=Coq,numbers=none,caption=Snippet of Coq script of \lstinline|ct04.v|, label=lst:ct04,frame=single]
Require Export Arith List.
...
Theorem div_conq_split : forall (P:list A -> Type), 
    P nil 
    -> (forall (a:A), P(a::nil)) 
    -> (forall (ls:list A), P fst(split ls) -> P snd(split ls) -> P ls
    -> forall (l : list A), P l.
Proof.
...
\end{lstlisting}

\begin{lstlisting}[language=Solidity,numbers=none,caption=Pseudo-code example of \lstinline|ct04Cont|, label=lst:ct04_cont,frame=shadowbox]
contract ct04Cont is registry {
    ...
    file="0x0ct04";
    CoqVer="8.12";
    filetype="Tactic";
    imports=["0x0Arith","0x0List"];
    right_to_use=pay_to_use([("div_conq_split",x)]);
    ...
}
\end{lstlisting}

After which, $T$'s tactic, \lstinline[language=Coq]|div_conq_split|, instead of \lstinline[language=Coq]|induction| as used in \lstinline[language=Coq]|ct01.v|, can be used to initiate an alternative proof via \lstinline[language=Coq]|ct05.v| (Listing \ref{lst:ct05}). 


\begin{lstlisting}[language=Coq,numbers=none,caption=Coq script of \lstinline|ct05.v|, label=lst:ct05,frame=single]
Require Export ct00 ct02 ct04.

Conjecture sort_prog_one : forall a : nat, {l' : list nat | sorted l' /\ permutation l' [a]}.

Conjecture sort_prog_split : forall (ls l' l'0: list nat), sorted l'0 -> permutation l'0 (fst (split nat ls))
    -> sorted l' -> permutation l' (snd (split nat ls)) -> {l'1 : list nat | sorted l'1 /\ permutation l'1 ls}.

Lemma sort_prog : forall (l : list nat), 
  {l' : list nat | sorted l' /\ permutation l' l}.
Proof. div_conq_split. apply sort_base. apply sort_prog_one. intros; destruct H; destruct a;
destruct H0; destruct a; eapply sort_prog_split. exact H. eassumption. exact H0. eassumption. Qed.
\end{lstlisting}

\begin{figure}[ht]
\centering
\includegraphics[width=\linewidth]{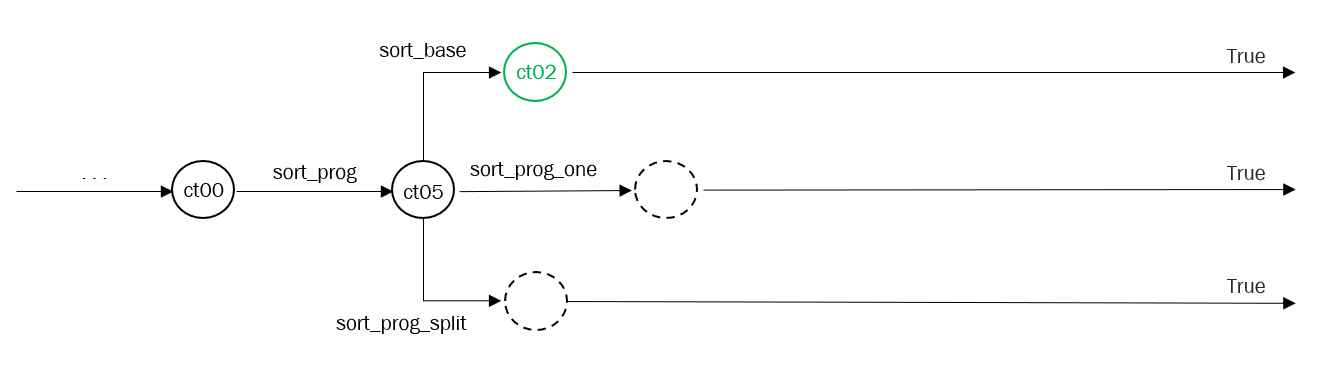}
\caption{New string diagram representation of the alternative proof presented in \lstinline[language=Coq]|ct05.v|. Note that \lstinline[language=Coq]|sort_base| is solved as $T$ imported and used the result from \lstinline[language=Coq]|ct02.v|.}
\label{fig:stringd_ct05}
\end{figure}

Once $T$ uploads the contribution of \lstinline[language=Coq]|ct05.v|
to the system, $C$ and the rest of the users will be notified of a new string diagram presentation of the alternative proof as shown in Fig. \ref{fig:stringd_ct05}.

\begin{figure}[ht]
\centering
\includegraphics[width=\linewidth]{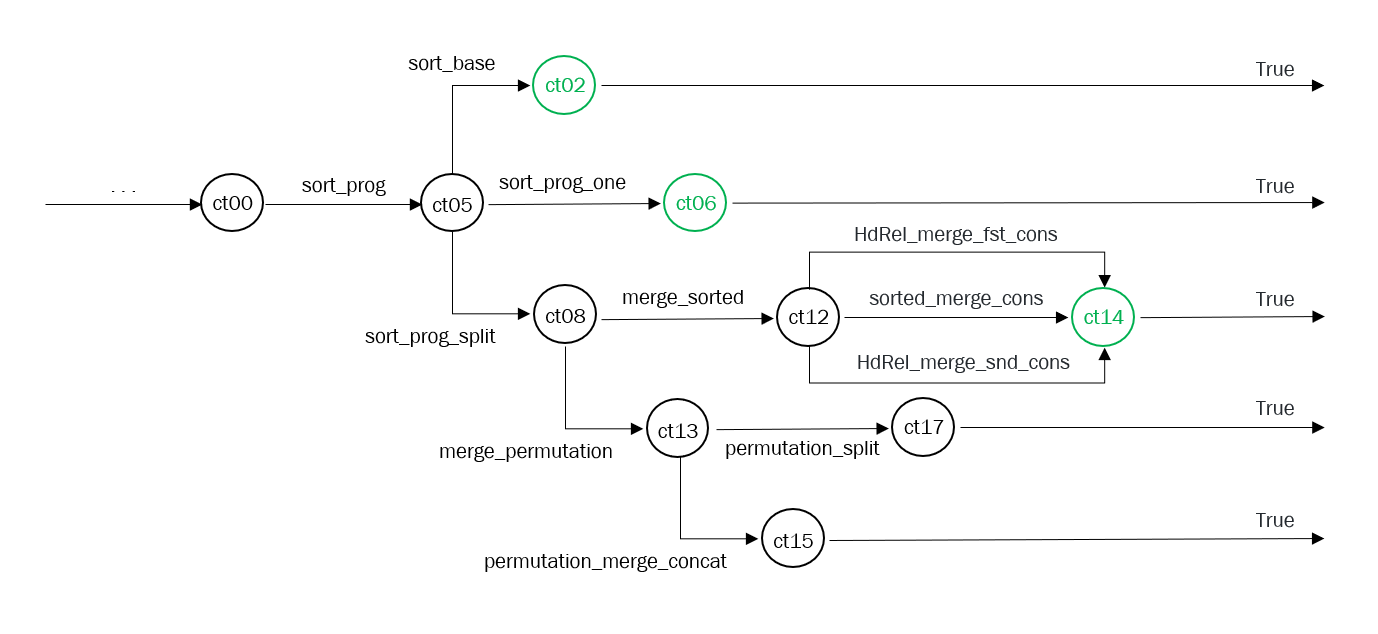}
\caption{Collaboration of contributions from human provers and automated AI tool/s that yields a proof of merge sort algorithm.}
\label{fig:stringd_msort}
\end{figure}

Suppose the insertion sort's proof (Fig. \ref{fig:stringd_isort}) is completed before the merge sort's proof. Once the latter proof is completed, the incentive mechanism will trigger and distribute \lstinline[language=Solidity]|TokenReward/2| appropriately to all the contributors in Fig. \ref{fig:stringd_msort} as it is the second completed proof.

\section{Conclusion and Future Directions} \label{sec:conclusion}

In our paper, we propose a blockchain-based approach to allow agents, human provers and automated AI systems, to formalize mathematics and programs collaboratively through the means of some proof assistants such as Coq. We broke down our system architecture into three dependent layers. The data, client and incentive layers describe what information is to be recorded, to be deemed as meaningful and to be rewarded respectively. We differentiate our work from the prior projects through i) the use of TCR to vote contributions in or out from the canonical registry, ii) smart contracts to implement incentive mechanisms to encourage both collaboration within the same proof and diversification of proofs, and iii) inclusion of automated AI tools. 
Moving forward, there are several directions we are looking to improve and decide upon before the  implementation of our proposal:

\subsection{Game theory of incentive mechanisms}

The flexibility of implementing complex incentive mechanisms via smart contracts allows us to reward the contributions within a proof and across different proofs appropriately. Thus, it is important to have the right incentive mechanism/s to encourage users to not only generate a single proof but also multiple proofs of a common problem collaboratively in a decentralized manner. Other than the incentive mechanisms mentioned in Section \ref{sec:example} \cite{banerjee2014re,babaioff2012bitcoin}, a game theoretical study on incentive mechanisms would provide us different ways to reward contributors appropriately.

It is natural to model the process of formalizing proofs collaboratively as a cooperative game where the contribution graph encodes the current state of knowledge and progress across a wide variety of open problems. Thus, it can be assigned a global value that can be captured natively in the system via publicly trade-able tokens.
In any such cooperative game, the overall system value can be distributed amongst its contributors using, e.g., the Shapley value distribution rule \cite{shapley1953value}. The Shapley value distribution rule has a number of well known and desirable axiomatic properties. For example, it is budget balanced\footnote{Sometimes this property is referred to as efficiency.}  meaning that the sum of the agents’ payoffs is equal to the global value of the system.
Moreover, it is the unique distribution rule that is efficient, symmetric, additive, and assigns zero payoffs to dummy players.
Finally, recent works have pointed out that within the set of budget balanced rules, it is effectively optimal \cite{dobzinski2018shapley,gkatzelis2016optimal,phillips2017design}.



\subsection{Usage of tokens} 

There can be two usages of the tokens generated for our blockchain. First, they can be used as a virtual currency purely to measure the contributions a user has made in the world of formalized mathematics and programs. This is analogous to the h-index \cite{hirsch2005index} to measure the productivity of a researcher's publications. Second, they can be exchanged for actual currencies, e.g., USD. This approach gamifies the process of formalization through monetary rewards, and hence, attracts more users to take part. The decision to use tokens as ``scores'' or as monetary incentive may depend on the community participating in our protocol.

\section*{Acknowledgments}
This research/project is supported in part by the National Research Foundation, Singapore under its AI Singapore Program (AISG Award No: AISG2-RP-2020-016), NRF 2018 Fellowship NRF-NRFF2018-07, NRF2019-NRF-ANR095 ALIAS grant, grant PIE-SGP-AI-2018-01, AME Programmatic Fund (Grant No. A20H6b0151) from the Agency for Science, Technology and Research (A*STAR) and the Ethereum Foundation. 

\bibliographystyle{unsrt}  
\bibliography{main}

\end{document}